# Multi-focal Picosecond laser vertical slicing of 6 inch 4H-SiC ingot

Jiabao Du[a,b,c,d], Shusen Zhao[a,b,c,d*], Xiaoyu Lu[a,b,c,d], Lu Jiang[a,b,c,d], Shifei Han[e,f,g], Xinyao Li[a,b,d], Xuechun Lin[a,b,c,d]

a Laboratory of All-solid-state Light Sources, Institute of Semiconductors, Chinese Academy of Sciences, Beijing 100083, China

b Center of Materials Science and Optoelectronics Engineering, University of Chinese Academy of Sciences, Beijing 100049, China

c College of Electronic, Electrical and communication Engineering, University of Chinese Academy of Sciences, Beijing 101408, China

d Engineering Technology Research Center of All-Solid-State Lasers Advanced Manufacturing, Beijing 100083, China

e State Key Laboratory of Precision Space-time Information Sensing Technology, Beijing, 100084, China

f Key Laboratory of Photonic Control Technology (Tsinghua University), Ministry of Education, Beijing，100084, China

g Department of Precision Instrument, Tsinghua University, Beijing，100084, China

* Corresponding author: E-mail address:.

Postal address: Laboratory of All-solid-state Light Sources, Institute of Semiconductors, Chinese Academy of Sciences, No. 35, Qinghua East Road, Haidian District, Beijing 100083, China.

**Abstract:** Ultrafast laser direct writing inside materials has garnered significant attention for its applications in techniques like two-photon polymerization, stealth dicing and vertical slicing. 4H-Silicon Carbide (4H-SiC) vertical slicing has wide potentials from research to industry due to low kerf loss and high slicing speed. In this paper, to improve the vertical slicing processing quality and lower the separation strength, we introduce a multi-focal vertical slicing method with spherical aberrations caused by refractive index eliminated. Additionally, by shaping the wavefront of picosecond laser, our experiments show the lower latitudinal ablation zone and higher crack propagation of multi-focal vertical slicing method on 4H-SiC, demonstrating that this method not only reduces filamentations to minimize ablation damage, but also significantly reduce the tensile strength during separation. We achieve a 6 inch 4H-SiC ingot vertical slicing and separation using 4-focal slicing. This method shows various potentials in laser processing.

## 1. Introduction

Ultrafast laser processing, including picosecond laser and femtosecond laser processing, has become an essential technique for precision material modification, particularly in specialized applications like stealth dicing and vertical slicing[1-3]. These applications are using the high peak intensity of ultrashort pulse to achieve minimum heat induced damage along the laser propagation area due to less total energy required and high center intensity at the focal point due to a series of ultrafast laser-matter interaction[4-6]. Among various materials, 4H-Silicon Carbide (4H-SiC), a third-generation semiconductor, demonstrates substantial potential due to its remarkable properties such as high thermal conductivity, chemical stability, and broad bandgap. Consequently, achieving efficient and minimal-damage vertical slicing of 4H-SiC is imperative for enhancing its applications in electronic and optoelectronic devices.[7-9]

Picosecond laser vertical slicing is to focus the laser inside 4H-SiC , the laser cause ablation zone inside the material, this damage includes laser ablation, shockwave induced damage [10] and the filamentation caused by the self-focus effect and plasma defocus effect [3, 11] which severely form a high longitudinal damage area causing high kerf loss[12, 13]. The cracks will induce a crack propagation alone the crystal cleavage plane.[14, 15] The laser ablation and shockwave induced damage are caused by laser pulse width, materials properties and laser intensity.[16, 17] Infrared picosecond laser is used in vertical slicing of 4H-SiC due to its high transmittance and high peak power.[12, 13] As reported, the laser filamentation is caused by the plasma defocusing and self-focusing due to the high peak intensity of the picosecond laser[18], another reason for filamentation is the spherical aberration caused by the mismatched refractive index of the air and 4H-SiC.[9, 19, 20] Thus, to achieve efficient vertical slicing with minimum kerf loss is to minimize the longitudinal damage and increasing vertical crack length.

Recent technological advances have incorporated spatial light modulator (SLM) into laser processing systems.[21, 22] SLM generate corrections phase for spherical aberrations and enable from multi-focal generation[23] to pattern printing[24], enhancing processing efficiency and precision.[25] Meanwhile, for stealth dicing, multi-focal[26], longitudinal focal length enlarging[27] and pseud-Bessel laser[28] can boost the processing efficiency and precision. Vertical slicing is the technology likely to stealth dicing but the cracks inside material is perpendicular to the laser propagation direction. Thus, the multi-focal perpendicular to the laser could increase the vertical slicing efficiency. However, traditional iterative methods for computer-generated holography (CGH) consume more time for high-speed manufacturing environments.[29-31] Researchers introduces fast iteration method like nonlinear point to point and use it into stealth dicing to improve the dicing quality[32] and is suitable for dynamic change in stealth dicing.

This study enhances the non-iterative multi-focal generation method initially proposed by Zhu et al.[33, 34], tailored for laser vertical slicing of 4H-SiC. By integrating spherical aberration correction, our approach aims to eliminate aberrations due to refractive index mismatches, thus improving the quality and uniformity of the focal points. Through vectorial diffraction simulations and experimental validation, we have demonstrated that our method can significantly reduce surface roughness and damage zones in 4H-SiC. These results underscore the potential of our non-iterative holographic technique to revolutionize ultrafast laser processing, providing a practical pathway for efficient and precise vertical slicing of transparent materials like 4H-SiC. Using this method, we achieved the vertical slicing of 6-inch 4H-SiC ingot.

## 2. Method

### 2.1 Vectorial diffraction modeling

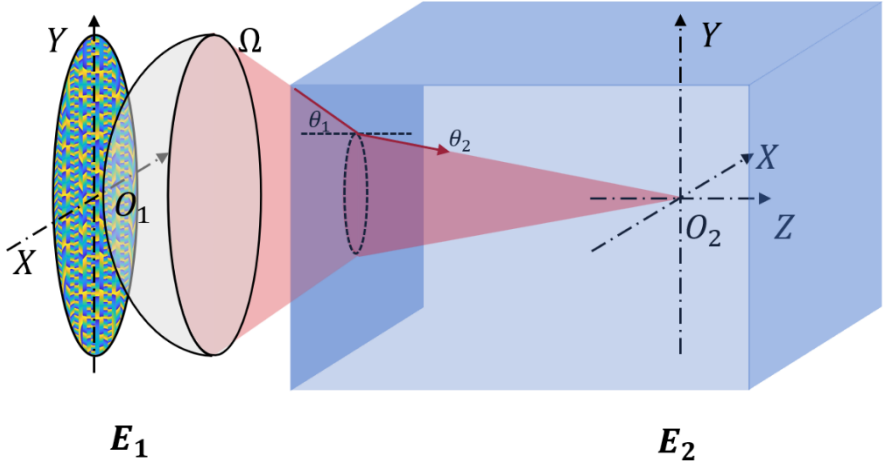

Fig.1 The principle of vectorial diffraction model and the scheme of spherical aberration caused by the mismatched refractive index

As shown in Fig.1, in high numerical aperture (NA) scenarios, the electric field distribution at the focal points is based on the Richard-Wolf vectorial diffraction theory[35]. By taking the refractive index mismatched situation, as shown in Fig.1, the electric field vector $\boldsymbol{E_2}$ in the focal region inside the material is expressed as follows:

$$\boldsymbol{E_2} = -\frac{i}{2\pi n_1 k_0} \boldsymbol{F}\{T(\theta_1, \theta_2, \phi)\boldsymbol{E_1} e^{ik_0 \varphi_{SA}} e^{ik_z z} \cos(\theta_1)\} \qquad (1)$$

Here, $\boldsymbol{E_1}$ is the electric field vector at the incident pupil. The matrix $T$ represents the influence of the objective lens $\Omega$ on different polarized incident components. In our experiments, the incident beam is linearly polarized, simplifying $\boldsymbol{E_1}$ to a single component. $d_{\text{nom}}$ is the nominal depth of the focal points which equals to the objective lens movement along z-axis, $\varphi_{SA} = d_{\text{nom}}(n_2 \cos\theta_2 - n_1 \cos\theta_1)$ represents the spherical aberration (SA) phase introduced by the material, $k_z z = k_2 z \cos\theta_2$ describes the intensity distribution as the beam propagates along the z-axis, $k_0 = 2\pi/\lambda$ is the wavenumber and $k_2$ is the wavenumber inside material, $n_1$ and $n_2$ are the refractive of air and 4H-SiC, respectively.

The spherical aberration phase, $\varphi_{SA}$, will cause the energy distribution loss along the z-axis, leading the low vertical slicing efficiency and larger longitudinal laser ablation zone. Meanwhile, the energy uniformity of each focal point inside the material will cause the different ablation depth which will cause the higher kerf loss. The relationship between the $E_1$ and $E_2$ is the Fourier transformation relation as described in formula (1). The CGH added onto the exponential term in $E_1$ within the Fourier transformation relation should form multi uniform focal points in $E_2$, as well as eliminating the spherical aberration included by the SA terms.

## 2.2 CGH generation method

We designed a non-iteration way to generate multi-focal inside material with spherical aberration correction. As shown in Fig.2 (a), the non-iteration is designed as two steps. First, the phase of focus shift is designed as: $\phi_S(x', y') = \frac{2\pi NA}{\lambda} \rho_{m,n} \Delta\rho_{m,n}$, the shift is alone the x direction. By applying the segmentation function into the phase shift function which is shown as $\Delta\rho(M, N)$, the focus can be shift by designed equally. Second, we applied the spherical aberration correction term $\phi_{SAC}(x', y')$ as eq. (2).

$$\phi_{SAC}(x', y') = \left(\frac{1}{s}\sqrt{\frac{n_2^2}{NA^2} - \rho_{m,n}^2} - \sqrt{\frac{n_1^2}{NA^2} - \rho_{m,n}^2}\right) d_{\text{nom}} \qquad (2)$$

The SA phase term can be eliminated by substituting this term on the exponential part, and the focal position can be shifted into the designed depth. By taking the two term of focal point separation, the focal shift phase is described as:

$$\phi(x', y') = \sum_{n=1}^{N} \sum_{m=1}^{M} \frac{2\pi NA}{\lambda} \left[\rho_{m,n}\Delta\rho_{m,n} - \left(\frac{1}{s}\sqrt{\frac{n_2^2}{NA^2} - \rho_{m,n}^2} - \sqrt{\frac{n_1^2}{NA^2} - \rho_{m,n}^2}\right) d_{\text{nom}}\right] \qquad (3)$$

Where, $M$ and $N$ are the number of rows and the number of focal shifts, $\Delta\rho_{m,n}$ is the focal shift of the

$M$-th and $N$-th row, we only need to set different values for $M$ and $N$ to represent the desired $N$ focal points and $M$ assignments, respectively, to achieve a uniformly distributed multi-focal point inside material. As shown in Fig.2 (a), $\Delta\rho(M,N)$ is a variable assigned to $\phi_S(x',y')$, and then add $\phi_{SAC}(x',y')$ onto the phase.

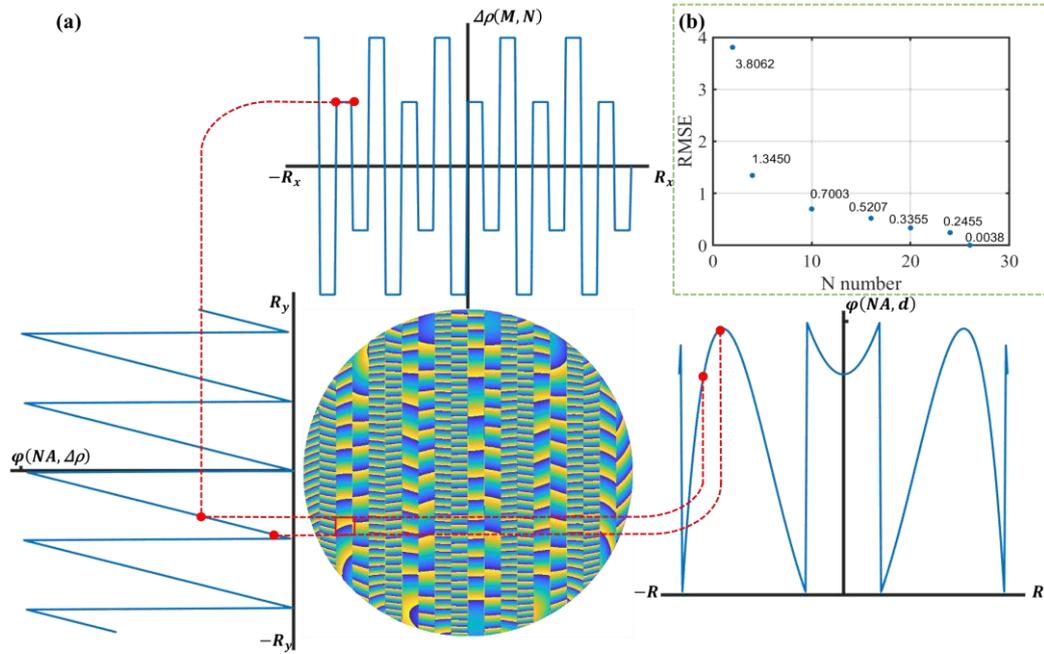

Fig.2

The principle of CGH generation

The energy distribution within the material can be calculated by taking the designed phase into the vectorial diffraction. we use the root mean square error $RMSE$ to analysis the quality of the multi-focal, which is defined as:

$$RMSE = \sqrt{\frac{1}{N}\sum_{i=1}^{N}(E_i - \hat{E}_N)^2}$$

Where, $E_i$ and $\hat{E}_N$ refers to the energy on $i$-th focus and $N$-th focus. The more approach to 0 of the $RMSE$ shows the uniform of the multi-focal, as shown in Fig.2 (b), the $RMSE$ decrease with the increase of the $N$ number. Here we use 4-focal as an example to show the $RMSE$ result, the uniformity of 4-focal achieve 0.0038 when *N* is larger than 26,comparing *RMSE*=0.01 of using Weighted Gerchberg-Saxton method for more than one minutes.[29] Based on the experiment, the pixel number is 832 based on the experiment designed later.

## 3. Experiment Setup

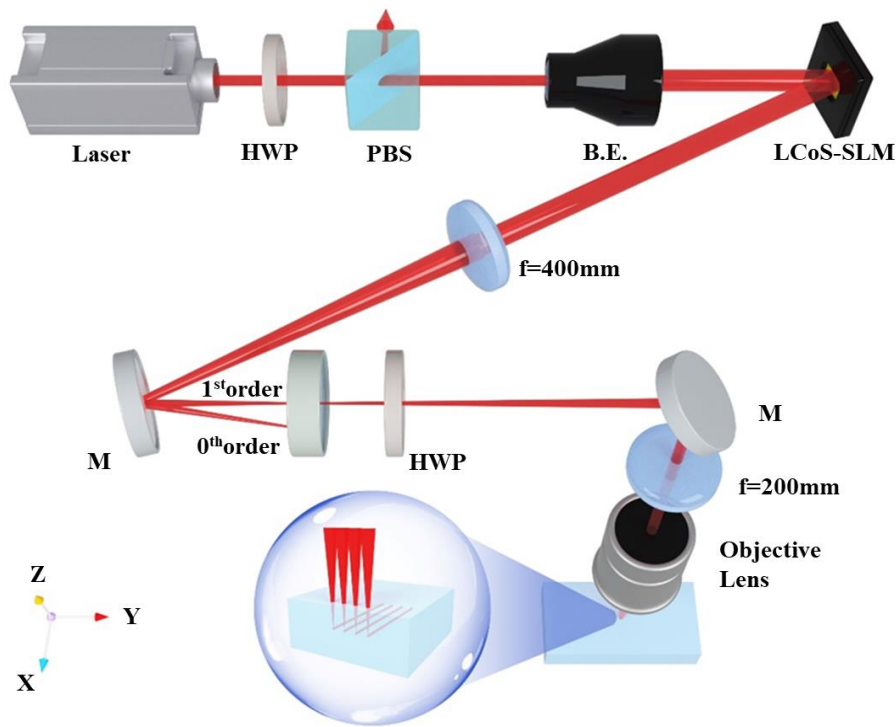

Fig. 3 Experiment setup

As depicted in Fig.3, the experimental setup utilizes a custom-built 3-axis motion stage, with the wafer secured by vacuum suction. We employed a laser sourced from Spectral Physics (Spirit HE 1040-30), characterized by a 12.6 picosecond pulse duration and a 1045 nm wavelength. The laser beam traverses a half-wave plate (HWP) and a polarization beam splitter (PBS) to control the single pulse energy and generate x-polarized light. This light is further expanded using a beam expander (B.E., LZBE-1030-1-4X, LBTEK) modulated by an LCoS-SLM (X15213, Hamamatsu). Adding the blazed grating phase on the SLM, the zeroth order diffraction beam is eliminated using a pinhole (P.H.). To control the polarization vertical to the scanning direction, another HWP is placed after pinhole. Our 4f optical system, featuring 400mm and 200mm focal lengths, achieves a 2:1 beam reduction before focusing through a 0.65 NA objective lens (M Plan Apo NIP HR 50X, Mitutoyo) inside the wafer.

The 4H-SiC wafer is sliced into 10mm × 10mm ×1.5mm and polished before experiment. By adjusting the z-axis, the laser focal depth is set to 500μm below the wafer surface, accounting for the mismatched refractive index. We designed various holograms onto the SLM, scanning the wafer along the x-axis to create inscribed lines with the same single pulse energy, repetition frequency, and pulse width. The wafer's side surface is then polished with sandpaper and examined under optical microscopes (Scope A1, Zeiss).

After selecting the optimal focal hatch, holograms of different focal number are employed to form a modified layer within the wafer. The wafer is affixed onto jigs and separated using a universal tensile machine (LTD CMT-30). After the separation, the vertical sliced wafer surface is analyzed using an optical microscope, scanning electron microscopy (SEM SU8020, Hitachi) equipped with an energy dispersive spectroscopy (EDS XFlash 6), and white light interferometry (Contour GT, Bruker).

## 4. Result and Discussion

### 4.1 Single scanning result

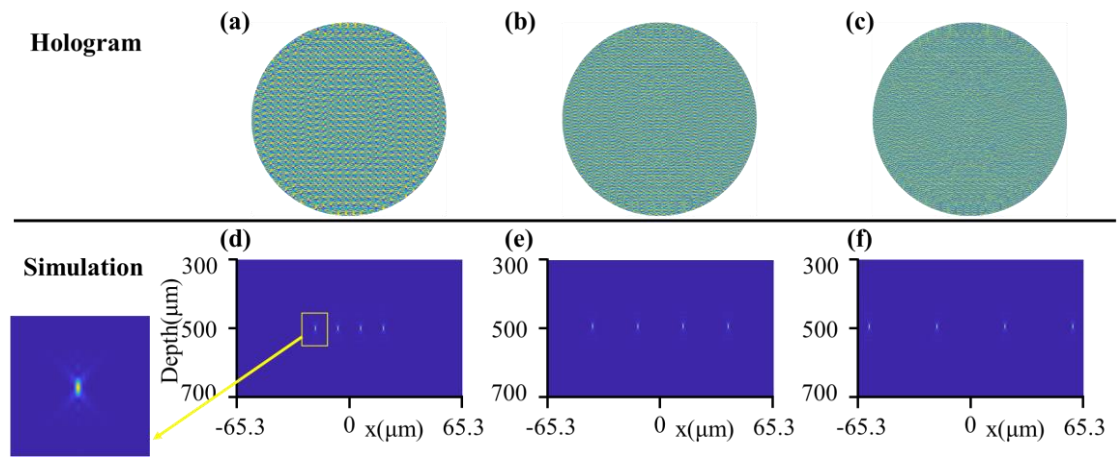

Fig.4 The simulation using holograms with 3 different focal hatches, (a)-(c): three different holograms referring to 20.4μm, 40.8μm and 61.2μm, (d)-(f): the simulation in vectorial diffraction model

Based on previous research[13, 20, 36-39], the interaction between picosecond laser pulses and 4H-SiC involves two primary mechanisms that contribute to cracking effects. Initially, cracks are observed in regions of high laser energy deposition, where significant ablation zones form due to nonlinear absorption coupled with heat transfer processes. These cracks propagate along the large ablation zones, the larger crack area results in less tensile strength, lower surface roughness and lower kerf loss. Additionally, for the high NA scenario, the focal point is around 2μm large, along with short pulse duration to form high peak energy density which surpass the nonlinear effect threshold for 4H-SiC to form filamentation effect, leading a longer longitudinal ablation.[40]

Dividing a single focal point into four focal points could improve crack propagation, because the diminished energy at each focal may shift the ablation zone smaller at the focal region. Moreover, the reduced energy is likely inadequate for generating extensive plasma channels, thereby minimizing overall kerf loss.

As illustrated in Fig.4(a)-(c), we designed three distinct holograms based on this method, with configurations of [-75λ, -25λ, 25λ, 75λ], [-150λ, -50λ, 50λ, 150λ], and [-225λ, -150λ, 150λ, 225λ], corresponding to 4-focal hatch of 20.4μm, 40.8μm, and 61.2μm, respectively. The laser energy distribution simulation of the hologram is shown in Fig.4(d)-(f). The energy distribution demonstrates the uniformity of the 4-focal point and the expand figure is shown, the laser is focused 500μm below the surface and effectively compensating for refractive index mismatches between 4H-SiC and air.

Upon applying the holograms to the SLM and adjusting the z-axis to position the laser focal point to be 500μm below the top surface (0001) of 4H-SiC. We set the single pulse energy at 25μJ with a frequency of 100 kHz, moving in the [11$\bar{2}$0] direction as indicated by prior research.[41] During the scanning process, the ablation zones are formed and the desired situation is the crack propagate between ablation zone.

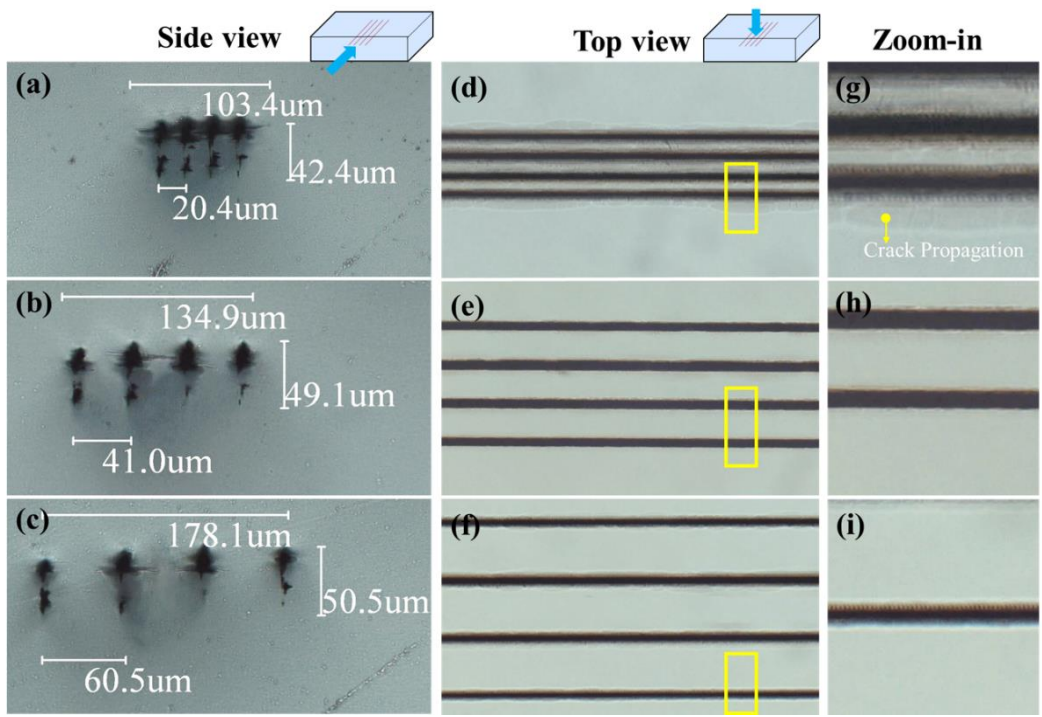

Fig.5 The single scanning test using holograms with 3 different focal hatches, (a)-(c): the side view of the 4H-SiC wafer after polishing;(j)-(l): the top view of the scanning path.

The side surface of the 4H-SiC, perpendicular to the scanning direction, as depicted in Fig.5(a)-(c). The observed hatch between the 4-focal agree with our simulation results. However, the side monotropy deviates from these results primarily due to ablation effects and nonlinear propagation over long distances. Ablation typically occurs around the geometric center of the focal point, creating ablation zone as black areas shown. These nonlinear effects extend these black areas into tips with a gap beyond the geometric center.

Fig.5(d)-(f) shows the crack propagation in the vertical direction along the ablation zone at shorter 4-focal hatch compared to larger hatch. The grey areas surrounding the black regions in Fig.4(d) indicate crack propagation, which is more likely to reduce stretching force and the kerf loss for vertical slicing of 4H-SiC. And only for shorter 4-focal hatch, the cracks propagate with each other due to the heat transfer around the focal point, which enable crack easier to propagate alone the heat affected area.

## 4.2 Vertical slicing of 4H-SiC

Based on prior experiment, the crack propagation between laser focal is evident when the separation of the 4-focal approximates 60μm. We utilized the phase patterns for 3 to 6-focal, consistently maintaining the total separation at 60μm. We operated a 3W laser at a repetition rate of 100 kHz and a scanning speed of 200 mm/s. The scanning hatch was set at 200μm along the [11-20] direction. the laser focal is adjusted to 500μm. below the material surface. At separation process, the wafer is adhered and subjected to tensile testing with a universal testing machine to measure the tensile force. After separation, the surface was analyzed for optical microscopy, surface roughness, SEM and EDS analyses.

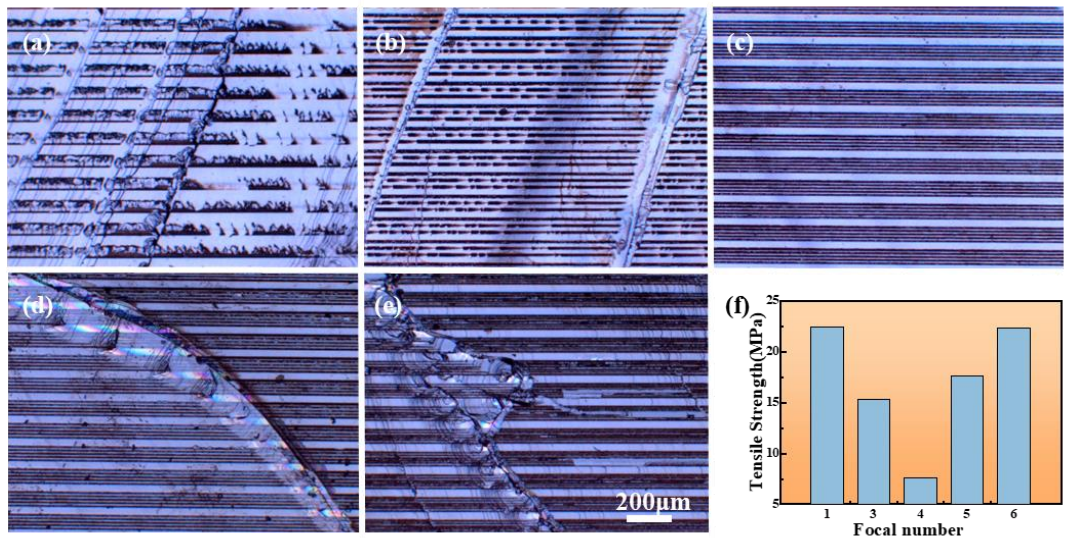

Fig.6 The experiment result of vertical slicing of 4H-SiC, (a)-(e) refers to the optical surface microscopy of single, 3, 4, 5 and 6-focal points laser slicing l, (f)refers to the tensile strength of the wafers during tensile test

Optical microscopy images of the vertical sliced wafer surface are shown in Fig.6(a)-(e), reveal that the surface monopoly is better compared to single focal vertical slicing comparing to the residual 4H-SiC on the vertical sliced surface. As the number of focal point increases, the energy at each focus diminishes. Optimal results are observed with 4-focal, which minimizes tensile strength due to more uniform crack propagation, single and 3-focal configurations result in less uniform crack connections and increased tensile strength. Increasing the focal number to 5 and 6 maintains crack propagation within a 60μm range but leads to poorer crack propagation due to reduced energy at each focus, and smaller heat affected zone as evidenced in Fig.6(f).

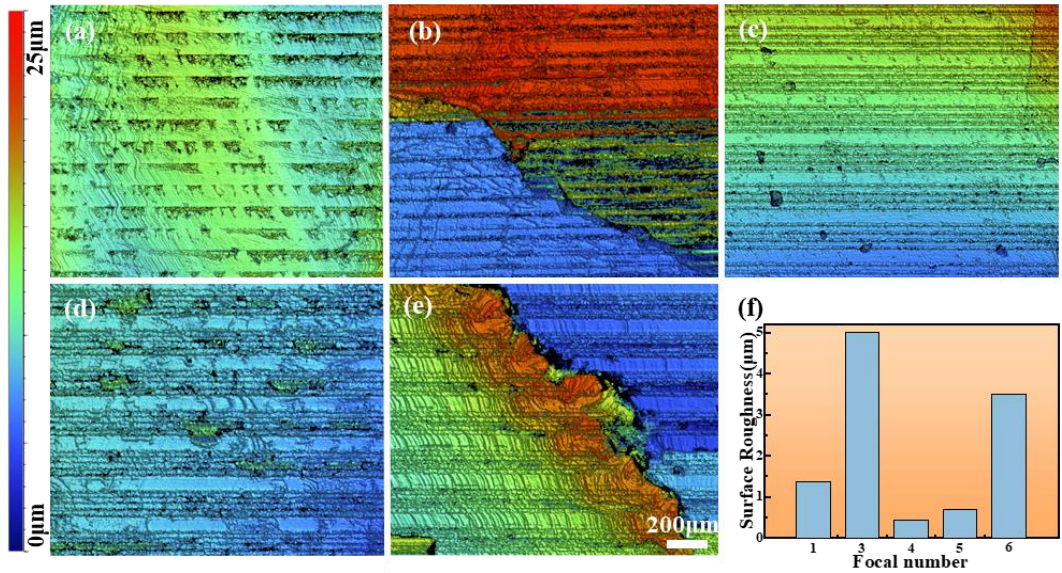

Fig.7 Surface roughness of vertical sliced wafer, (a)-(e) shows the surface of separated wafer and color bar is 0μm from blue to 25μm of red, (f)shows the total surface roughness (Ra) of 5 wafers.

Surface roughness measurements via white light interferometry, as depicted in Fig.7(a)-(e), indicate that surface roughness decreases initially and then increases as the focal number increase. With more focal, the ablation zone is less and the energy of each focal reduce causing weaker filamentation effect, enhancing crack connections and smoothing the post-tensile surface. However, an increase in the focal number leads to insufficient energy to effectively connect ablation zone by crack propagation between scans spaced 200 μm, resulting in a less uniform final vertical sliced surface. For 4-focal scanned surface, the surface roughness is only 432nm.

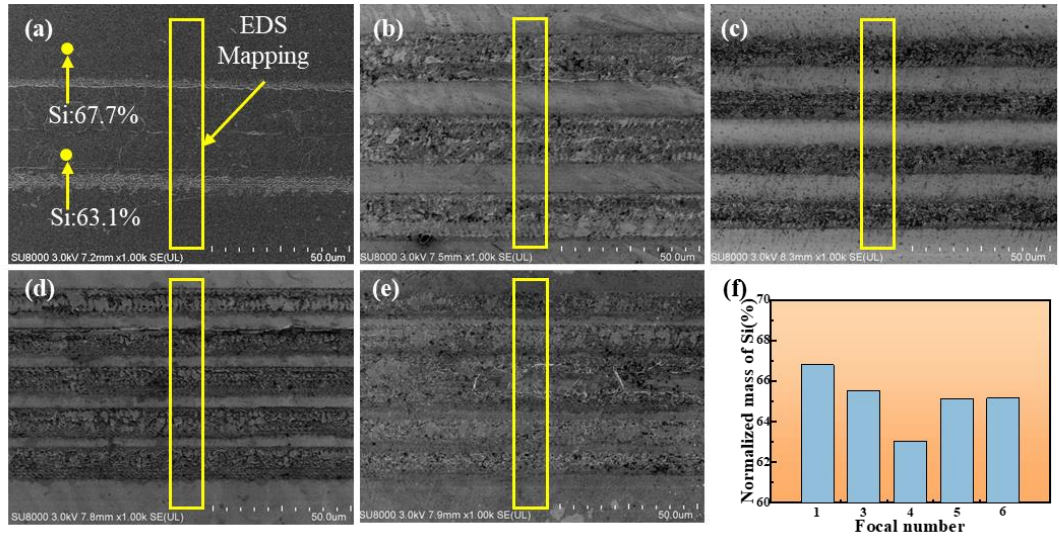

Fig.8 SEM and EDS test on separated wafers, (a)-(e) is the SEM image of separated wafer surfaces by single, 3, 4, 5 and 6-focal points laser slicing, yellow box refers to the EDS analysis scanning line and (f) refers to the EDS result of Si components

Further, SEM and EDS analyses of the vertical sliced wafer surface, presented in Fig.8(a)-(e), demonstrate that t the multi focal vertical slicing can increase a higher energy concentration in the focal depth. The vertical sliced wafer surface is combined with Silicon (Si) and Carbon (C) and Oxygen(O), as the ablation temperature increases, Si transforms into amorphous Si, while the higher melting point of C results in a higher residual content. As shown in Fig.8 (a), the Si component in the ablation area is less than that in the crack area, which agrees with the discussion above. Furthermore, the EDS mapping a tested as the yellow box area shown in Fig.8 (a)-(e). Fig.8(f) shows that using 4-focal results in a Si content of only 63.03% in the ablation zone, indicating a larger ablation area and higher temperatures at the same cross-section, thus yielding a smoother surface. With fewer than 4-focal, the high nonlinear threshold per focus leads to large longitudinal energy distribution through filamentation. Conversely, using more than 4-focal reduces the single focal energy, decreasing the heat affected zone size and increasing the Si component after vertical slicing.

From these observations and analyses, we conclude that effective vertical slicing of 4H-SiC contains two process of laser ablation causing heat affected zone and the crack propagation connected with each ablation zone. In this experiment, the multi-focal will decrease the single focal point energy and reduce the filamentation effect, also will enhance the crack propagation due to the vertical multi-focal distribution. We found for the specific energy and repetition frequency, the 4-focal vertical slicing method will reduce the separation tensile strength, surface roughness and reveal that the Si component will reduce due to more heat affected zone in the cross section.

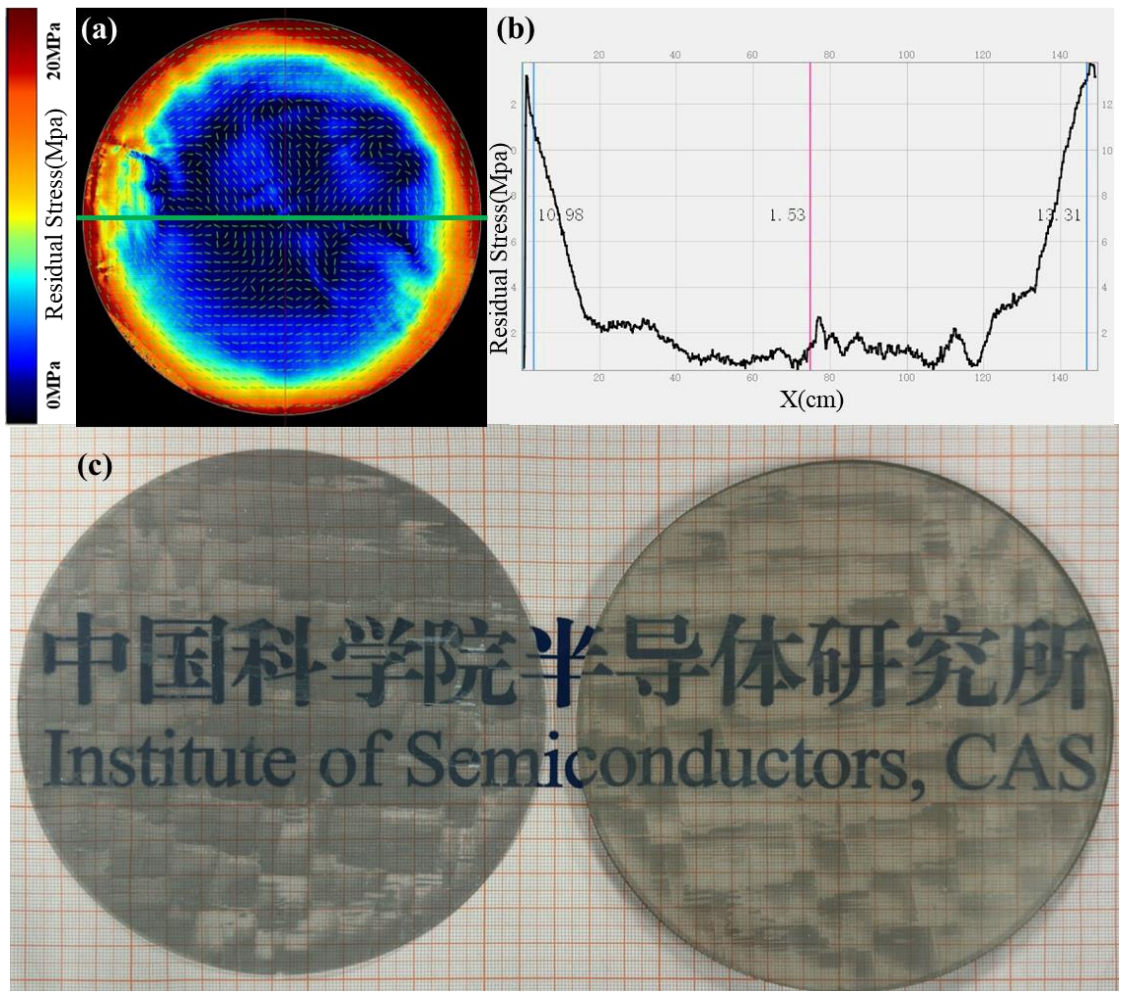

Fig.9 (a)-(b) Residual Stress test result on scanned 4H-SiC ingot and (c) Wafer and ingot after separation

Based on this experiment, we realized a 6-inch 4H-SiC ingot vertical slicing. After scanning, the residual stress at the ingot is tested. Because the laser for a crack initiation and a propagation, the larger residual stress after scanning means the more crack under the surface of the ingot, which means the crack will be easier to elongate with each other. Fig.9(b) is the residual stress distribution of green line in Fig8.(a). As shown in Fig.9(a)-(b), the residual stress shows the higher residual stress is around the side of the scanned ingot, the maximum residual stress is 15Mpa, which means the outer area of the scanned ingot is easier to separate after scanning because the residual stress indicates the crack initiation around the ingot, the ingot will separate from the higher residual stress area into the low stress area. The trend line in Fig.9(a) is perpendicular to the direction of stress distribution. This is because we have multiple scans that are perpendicular to each other during the scanning process, resulting in a stress distribution that appears to radiate from the outside inwards. As The crack initiation is around the ingot and when applying force onto different surface, at separation process, the 500μm wafer will separate along the crack propagation. The crack will propagate with each other from the outer side to the inner side on the ingot. After the separation process, the ingot was tested to be no residual stress left and the separated wafer and ingot are as shown in Fig.9.(b).

## 5. Conclusions

In this study, we developed a method for hologram generation in laser processing inside the 4H-SiC wafers. This non-iterative approach demonstrates significant potential for rapid calculation of hologram. The experimental results show that the method results in reduced damage compared to traditional spherical aberration correction phases. Analysis indicates that the 4-focal method for 3W laser power effectively minimizes the damage zone during laser vertical slicing of 4H-SiC by reducing large area ablation and nonlinear effects. EDS test results show more Si component on the separated surface 4-focal after vertical slicing, which indicates higher heat concentration in the focal plane compared with more or less focal-number due to the filamentation induced damage by less focal number or the lower focal energy. Furthermore, this method facilitates smoother surfaces, a decrease in surface roughness from 1.3 μm to 432 nm using the 4-focal, and the tensile strength reduced from 22.4MPa to 7.6MPa. Using this method, we realize a 6-inch 4H-SiC ingot vertical slicing and separation, the maximum residual stress on the scanned surface is 15Mpa and shows a higher residual stress from outer to inner of the ingot scanned area. This method shows a potential use for ultrafast laser processing inside the material and laser direct writing.

**Figure Captions：**

Fig.1. The principle of vectorial diffraction model and the scheme of the spherical aberration caused by the mismatched refractive index.

Fig.2. The principle of CGH generation.

Fig.3. Experiment setup.

Fig.4. The single scanning test using holograms with 3 different focal gap, (a)-(c): three different holograms referring to 20.4μm, 40.8μm and 61.2μm, (d)-(f):the simulation in vectorial diffraction model, (g)-(i): the side view of the 4H-SiC wafer after polishing;(j)-(l): the top view of the scanning .

Fig.5. The experiment result of vertical slicing of 4H-SiC, (a)-(e) refers to the optical surface structure of 1 3 4 5 6 focus, (f)refers to the tensile strength of the wafers during tensile test

Fig.7. Surface roughness of sliced wafer, (a)-(e) shows the surface of separated wafer and color bar is 0μm from blue to 25μm of red, (f)shows the total surface roughness (Ra) of 5 wafers.

Fig.8 SEM and EDS test on separated wafers, (a)-(e) is the SEM image of separated wafers surface1 3 4 5 6 focus laser slicing, yellow line refers to the EDS analysis scanning line and (f) refers to the EDS result of Si components

Fig.9. (a) Residual Stress test result on scanned 4H-SiC ingot and (b) wafer and ingot after separation.

**Declarations**

**Availability of data and materials**

The authors declare that the data supporting the findings of this study are available within the paper。

**Competing interests:**

The authors declare that they have no competing interests.

**Funding:** This work was financially supported by National Key R&D Program of China (No.2022YFB3605800), the CAS Project for Young Scientists in Basic Research, Grant No. YSBR-065, National Natural Science Foundation of China (No.62225507, No. U2033211, No.62175230, No.62175232, No.62275244), Scientific Instrument Developing Project of the Chinese Academy of Sciences, Grant No. YJKYYQ20200001, Key Program of the Chinese Academy of Sciences (ZDBS-ZRKJZ-TLC018), Beijing Municipal Science & Technology Commission, Administrative Commission of Zhongguancun Science Park (Z231100006023010).

**Authors' contributions: Jiabao Du**：Investigation, Writing original draft, Formal analysis, Investigation, Validation. **Shusen Zhao:** Conceptualization, Writing – review & editing, Supervision. **Xiaoyu Lu:** Methodology, Data analysis, Discussion. **Lu Jiang:** Discussion, Data analysis. **Shifei Han:** Investigation, Formal analysis. **Xinyao Li:** Investigation, Writing. **Xuechun Lin:** Supervision, Funding.

**Acknowledgements**

Useful suggestions and revisions of the English version of this manuscript given by Jingyuan Zhang are also acknowledged.